\documentstyle[floats,twocolumn,amssymb,aps,epsfig,pre]{revtex}
\draft

\begin{document}
\wideabs{ \title{Stress transmission through three-dimensional ordered
granular arrays} \author{Nathan W. Mueggenburg,
Heinrich M. Jaeger, Sidney R. Nagel} \address{The James Franck
Institute and Department of Physics\\ The University of Chicago \\
5640 S. Ellis Ave. Chicago, IL 60637} \date \today \maketitle

\begin{abstract}
We measure the local contact forces at both the top and bottom
boundaries of three-dimensional face-centered-cubic and
hexagonal-close-packed granular crystals in response to an external
force applied to a small area at the top surface.  Depending on the
crystal structure, we find markedly different results which can be
understood in terms of force balance considerations in the specific
geometry of the crystal.  Small amounts of disorder are found to
create additional structure at both the top and bottom surfaces.
\end{abstract}

\pacs{PACS numbers:  81.05.Rm, 45.70.-n, 83.70.Fn, 45.70.Cc}
}

\section{Introduction}

In a static granular bead pack all of the forces on a particle, due to
the contacts from its neighbors and due to gravity, must be in perfect
balance; there can be no net force on any particle.  Moreover, in the
limit of hard particles, no particle distortion is allowed leading
several investigators to question whether elasto-plasticity theory can
be used in this hard-sphere limit to describe the mechanical
properties of an amorphous granular material
\cite{Bouchaud95,Wittmer96,Edwards98,Head01}.  Others suggest that
elastic theory is capable of producing the observed experimental
results \cite{Goldenberg01,Goldhirsch02}.  The presence of disorder
makes an exact calculation of the contact forces in such a material
impossible.  On the other hand, it should be possible to calculate the
forces exactly in a perfect crystal \cite{Stott98}.  However, because
macroscopic grains do not spontaneously crystallize, it has been
difficult to create large defect-free granular crystals
\cite{Pouliquen97,Vanel97,Scott64,Berg69} on which to perform an
experiment to see how the forces are transmitted.  In addition, there
will always be small imperfections or asperities in the particles
themselves and slight variations in the diameters of the grains.
Although we cannot completely avoid disorder of this latter kind, we
have found an efficient method of constructing essentially perfect
crystals with few, if any, defects in the particle positions
\cite{Blair01}.  With such crystals, one would hope that the presence
of small disorder from the variation of the individual particle shapes
and sizes could be treated as a perturbation about the perfect
crystalline response.

Previously, there have been experimental studies of the statistical
properties of the local contact forces in granular packs.  In
particular, the distribution of normal forces, $P(F)$, between the
particles and the container walls was measured in a variety of
situations \cite{Blair01,Mueth98,Makse00,Lovoll99}.  In most of those
studies, a piston pressed down on the entire top surface of the
granular material.  The surprising result was that the entire form of
$P(F)$ (a large value near $F = 0$, a small peak [or plateau] with a
maximum near the average force and an exponential tail at large
forces) was robust and did not depend on whether the granular medium
was amorphous or crystalline or whether the particles were smooth or
rough \cite{Blair01}.  The question is thus raised as to what aspects
of the force propagation in a granular material are affected by
preparation and the underlying crystal or amorphous structure.

Several experiments have studied the response (i.e., the Green's
function) to a localized external force at the top surface of a
granular medium.  The different studies found somewhat contradictory
results.  DaSilva and Rajchenbach \cite{DaSilva} studied a system
consisting of a special two-dimensional packing of photoelastic
rectangular bricks.  They measured the contact forces between
particles as a function of position and found that the contact forces
had a maximum centered below the position of the applied perturbation
and that on ether side the forces decreased in magnitude. The width of
this peak in the forces grew as the square root of the depth below the
surface.  This result is consistent with models predicting a diffusive
propagation of the forces such as the q-model proposed by Coppersmith
{\em et al.} \cite{Liu95,Coppersmith96}.  Other studies have focused
on the response to a localized external force in amorphous packings of
spheres.  Reydellet and Cl\'ement found that the response at the
bottom of a three-dimensional pack had a central maximum of
approximately Lorenztian shape with a width that grew lineraly with
the depth of the pack \cite{Reydellet01}.  Similar behavior was found
in two-dimensional samples studied by Geng {\em et al.} \cite{Geng00}.
This linear increase of the width with depth is consistent with
elastic theories \cite{Goldenberg01,Goldhirsch02}. The work by Geng
{\em et al.} also suggests a strong effect due the spatial ordering of
the particles resulting in a central minimum of the response forces in
ordered packings.  Thus there is a need to study the propagation of
forces from a localized perturbation in a nearly perfectly ordered
crystalline array of particles in order to investigate how the
underlying order (or disorder) of the lattice influences the force
propagation.

Bouchaud {\em et al.} have proposed a model in which forces propagate
in straight lines until encountering a defect where they are split,
and they have shown that this model leads to a central force minimum
for small depths and to elastic-like behavior for large pack depths
\cite{Bouchaud00}.  The straight-line propagation of forces (resulting
from the hyperbolic model of Cates {\em et al.}  \cite{Cates98}) and
the splitting of forces can be viewed in a geometrical framework based
upon the contact orientations of neighboring beads.  From this
viewpoint crystalline packings allow for the control of these contact
orientations and the resulting splitting.  In particular, the
hexagonal-close-packed (hcp) crystal structure is such that forces are
required to split at nearly every layer as they travel downward
through the pack.  By looking at the response of an hcp crystal to a
localized force it is possible to explore the consequences of multiple
splittings in comparison with the predictions of the forcechain
splitting model.

In this paper we present an experimental investigation of the response
to a localized external force in three dimensional nearly perfect hcp
and face-centered-cubic (fcc) crystals.  We measure the resulting
contact forces at both the top and bottom surfaces of the crystal, and
investigate how the pattern of forces depends on the crystal structure
and the depth of the pack.  We determine that in the regime of highly
ordered granular packs, the crystal structure plays a dominant role in
determining the average pattern of response forces at the bottom
surface.  These patterns can be viewed as a consequence of force
balance in the specific geometry of the crystal.  We also consider the
effect of small amounts of disorder in these crystals and show that
they modify the pattern of forces on the bottom surface in a
nontrivial manner and scatter forces back upwards through the pack to
the top surface.

\section{Experimental Methods}

We created large nearly perfect hcp and fcc granular crystals of $3.06
\pm 0.04$ mm diameter soda lime glass spheres.  These crystals of
approximately $53$ bead diameters on a side were contained within an
acrylic cylinder with a triangular cross-section designed to be
commensurate with the hcp crystal structure.  The crystals were
created as individual layers of triangular order in the manner
described in reference \cite{Blair01}.  In this way the position of
each individual bead was viewed and corrected if neccessary as the
crystal was constructed, so that the resulting crystal would have very
few defects (estimated to be less than ten beads significantly out of
place for a crystal of approximately $50000$ beads).  Furthermore, the
stacking order of the planes was controlled in order to achieve three
dimensional fcc and hcp crystals.

After building such a crystal we applied an external force to a
localized region at the center of the top of the crystal either by
using a slowly applied steady force of approximately $100$ pounds or
by using a quick impulse force.  When using the slowly applied force
it was necessary to apply the force over a region approximately six
beads in diameter.  Attempts to apply the force to a smaller region
resulted in significant bead movement.  When using the quick impulse
force it was possible to apply the force to a smaller area of
approximately two beads in diameter.  For the purpose of comparison
experiments were also done with the quick impulse force applied to the
large six bead diameter region.

\begin{figure}[tb]
\centerline{\epsfxsize= 8.0cm\epsfbox{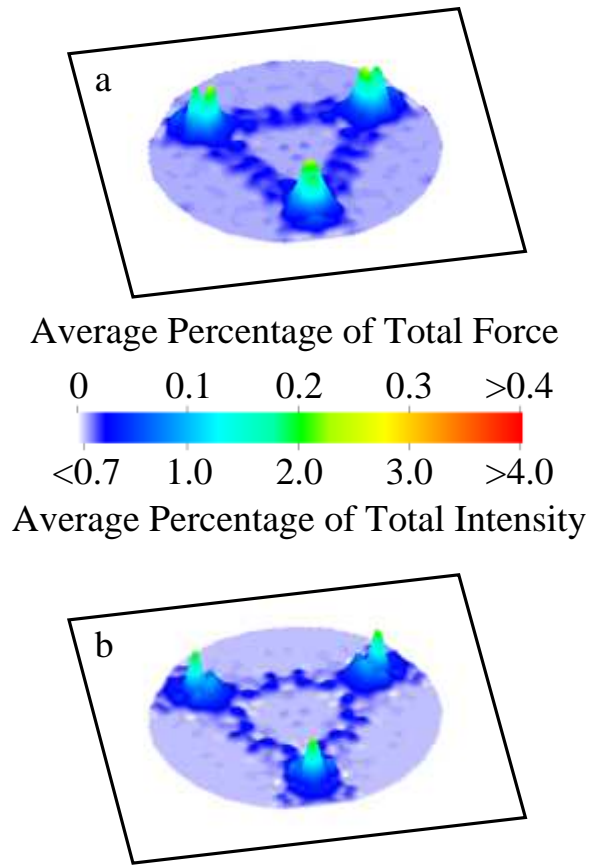}}
\vspace{.5cm}
\caption{(Color) A comparison of the pattern of calibrated forces in
response to a slowly applied force to the pattern of intensities in
response to a quick impulse force.  (a) The pattern of calibrated
forces at the bottom surface of a $19$ layer fcc crystal in response
to a slowly applied force over an area six beads in diameter at the
top surface.  (b) The pattern of intensities of carbon marks in
response to a quick impluse force applied to the same crystal
structure over the same area.  Both pictures represent an average over
ten experimental runs and an average over crystal symmetries.}
\label{force_intensity}
\end{figure}

We used a carbon paper technique to measure the forces in response to
the external forces \cite{Blair01,Mueth98,Liu95,Delyon90}.  A piece of
carbon paper and a piece of white paper \cite{materials} were placed
on the top and bottom of the crystal.  As the force was applied
individual beads pressed into the carbon paper and left marks on the
white paper. The size and darkness of the resulting marks was related
to the normal force on the corresponding beads.  When using the slowly
applied force, we calibrated the force versus darkness response as
explained in \cite{Blair01}.  The positions of the carbon marks were
matched to a triangular lattice and the forces for each lattice point
were averaged over approximately ten experimental runs.  It is
important to note that individual runs vary significantly, but that
the average over many runs remains highly reproducible.  In order to
improve statistics, the locally measured forces were also averaged
over symmetries of the crystal (one reflection and two rotations).

Note that the analysis technique must be carried out carefully so as
not to introduce error from the averaging over many files, and crystal
symmetries.  If the centers of the images are not determined
accurately, then lattice positions from file to file will not match up
exactly and will result in a small degree of broadening of any
features.  We estimate this broadening to be significantly less than
one bead diameter for each of the regions of large force.

When using the quick impulse, we calibrate the measured forces by
comparing them with those obtained with an identical set-up using the
slowly applied force described above.  We compared the patterns of
carbon marks left by the two methods and found no qualitative
difference.  Figure \ref{force_intensity}(a) shows the pattern of
calibrated forces at the bottom of a $19$ layer fcc crystal in
response to a slowly applied force at the top surface over a region
six beads in diameter.  The calibration maps points of zero intensity
to a small but nonzero force which results in a nonzero background.
Figure \ref{force_intensity}(b) shows the pattern of intensities of
carbon marks at the bottom of a similar crystal in response to a quick
impulse force applied to the same area.  The color scale of intensity
has been shifted by a factor of ten and a minimum value used to match
the previous pattern of forces.  Using this shift in color scale
intensity we map the two different measurements onto one another and
obtain a calibration for the quick impulse.

\section{Results}

In amorphous packings it is difficult to apply a localized force large
enough to leave carbon marks without also having significant movement
of the grains.  As such we were only able to explore the response of
amorphous packings to the quick impulse force applied to a large six
bead diameter region.  In this case we observed the pattern of
intensities at the bottom of three-dimensional amorphous packs of
various depths and found a broad central maximum.  Our data on the
width of this central maximum are consistent with a square-root growth
with depth as seen in a two-dimensional packing of bricks by DaSilva
and Rajchenbach \cite{DaSilva}, but are also within error of a linear
growth as seen by Reydellet and Cl\'ement \cite{Reydellet01} in
three-dimensional packings.

\subsection{FCC crystals}

\begin{figure}[tb]
\centerline{\epsfxsize= 8.0cm\epsfbox{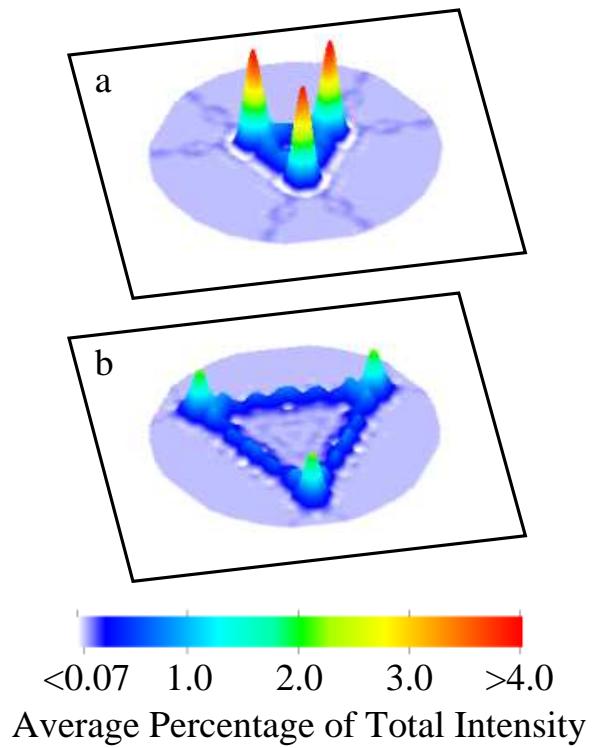}}
\vspace{0.5cm}
\caption{(Color) Intensity patterns at the bottom surface of fcc
crystals (a) $9$ and (b) $19$ layers in depth in response to a quick
impulse force applied to a two bead diameter region at the top surface.}
\label{fcc_surfaces}
\end{figure}

When using highly ordered packings bead movement is much less of a
problem, and it is possible to use a slowly applied force to a large
area or a quick impulse to a smaller area.  For fcc crystals we found
a characteristic response pattern at the bottom surface which is
qualitatively independent of type of force applied.  Figure
\ref{fcc_surfaces} shows the patterns of intensities at the bottom of
fcc crystals, $9$ layers and $19$ layers in depth, in response to a
quick impulse force applied to an area two beads in diameter at the
center of the top surface of the crystal averaged over approximately
ten experimental runs and averaged over crystal symmetries.  We find
three regions of large force, arranged in a triangular pattern.  The
size of this triangle increases with the depth of the crystal.  The
size of the three regions of large force are comparable to the size of
the region over which the force was applied.  In addition to the three
regions of large force, there appear fainter but still significant
regions of force along the lines connecting these three regions.

\begin{figure}[p]
\centerline{\epsfxsize= 8.0cm\epsfbox{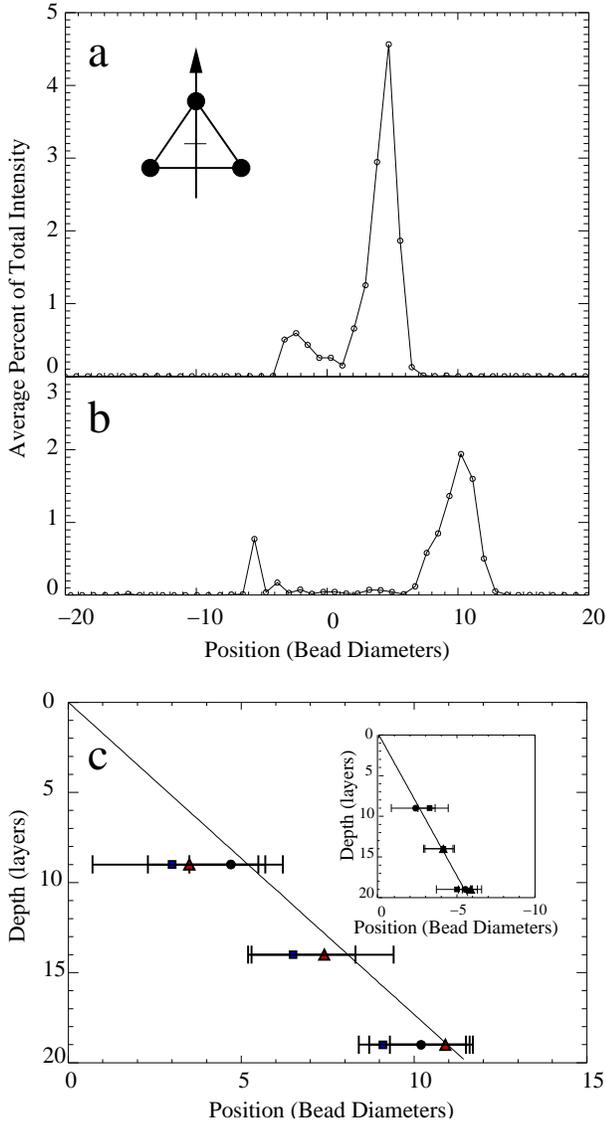}}
\vspace{.5cm}
\caption{Intensity as a function of position at the bottom of FCC
crystals (a) $9$ and (b) $19$ layers in depth in response to a quick
impulse force applied to an area two beads in diameter at the top of
the pack.  The cross-section was taken through the triangular pattern
of intensity as shown in the inset.  (c)  The position of main
intensity peaks as a function of the depth of the pack in response to
(circles) a quick impulse force over a small area, (triangles) a quick
impulse force over a large area, and (squares) a slowly applied force
over a large area.  Error bars represent the full width at half the
maximum value of the peaks. The black line indicates a slope of $35.3$
degrees with respect to the vertical which corresponds to an angle
specific to the geometry of the fcc crystal.  The inset shows the
position of the smaller peaks corresponding to the edge of the
triangular pattern with a line representing an angle of $19.5$
degrees.}
\label{fcc_profiles}
\end{figure}

Figure \ref{fcc_profiles}(a,b) shows cross sections of the intensity
versus position across the triangular pattern at the bottom surface of
$9$ and $19$ layer fcc crystals.  As depicted in the inset of figure
\ref{fcc_profiles}(a) the cross-section proceeds across the edge of
the triangular pattern, corresponding to a small peak, through the
center of the pattern (position $=0$) with small intensities, and
across the region of large force, corresponding to a large peak.  The
position of the large peaks as a function of the depth of the crystal
are indicated in figure \ref{fcc_profiles}(c), with the error bars
corresponding to the full width at half maximum of the peaks, which
can in some cases be seen to be significantly skewed.  Additional data
points are included in part c corresponding to experimental runs with
a quick impulse applied to a large area and to runs with a slowly
applied force over a large area.  The sloped line indicates an angle
of $35.3$ degrees which corresponds to the angle down the edge of a
tetrahedron. The regions of large force are found to lie just inside
of this angle as explained in section \ref{Interpretation}.  The inset
of figure \ref{fcc_profiles}(c) shows the position of the small peaks
corresponding to the edge of the triangular patterns as a function of
the depth of the crystal packing.  The solid line shows an angle of
$19.5$ degrees with respect to the vertical, which is corresponds to
the angle down the face of a tetrahedron.

\subsection{HCP crystals}

\begin{figure}[tb]
\centerline{\epsfxsize= 8.0cm\epsfbox{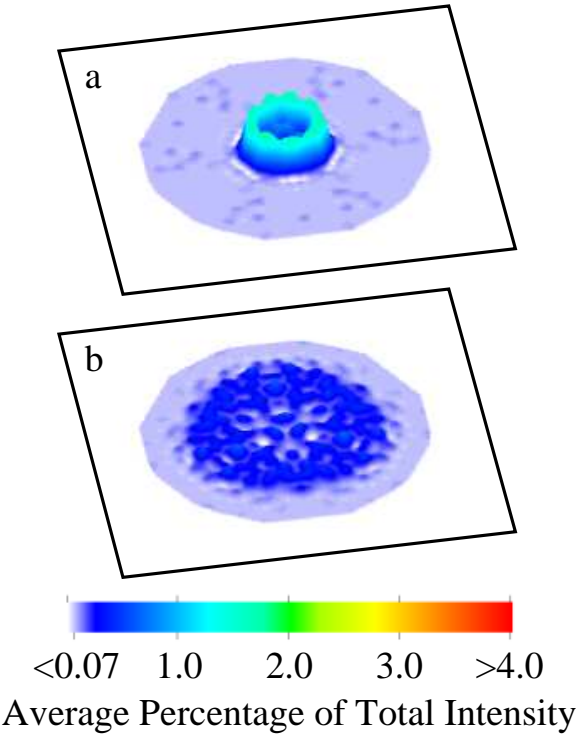}}
\vspace{.5cm}
\caption{(Color) Intensity patterns at the bottom of hcp crystals (a)
$9$ and (b) $25$ layers in depth in response to a quick impulse force
over an area two beads in diameter at the top surface of the crystal.}
\label{hcp_surfaces}
\end{figure}

\begin{figure}[htb]
\centerline{\epsfxsize= 7.0cm\epsfbox{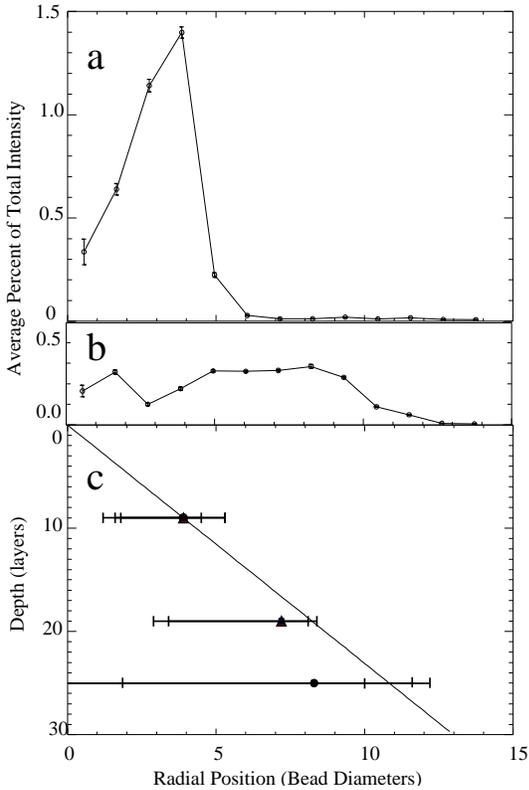}}
\vspace{.5cm}
\caption{Radial distribution of intensity at the bottom of hcp
crystals (a) $9$ and (b) $25$ layers in depth in response to a quick
impulse force applied to a small area at the top surface.  The
position of the peaks as a function of the depth of the pack are shown
in (c) as circles with error bars representing the full width at half
of the maximum value of the peak. As the interior of the ring is
nearly uniform no lower bound is shown for the $25$ layer crystal.
The triangular data points indicate similar peaks in response to a
quick impulse force over a large area, and the square data points
correspond to a slowly applied force over a large area. The solid line
corresponds to an angle of $27.4$ degrees with respect to the vertical
which is the average of two angles specific to an hcp crystal
structure as explained in the text.}
\label{hcp_rad_dist}
\end{figure}

In contrast to the triangular patterns found for fcc crystals, hcp
crystals show rings of large force at the bottom surface in response
to a localized force at the top surface.  Figure \ref{hcp_surfaces}
shows the intensity patterns at the bottom surfaces of hcp crystals
$9$ and $25$ layers in depth in response to a quick impulse force
applied to an area two beads in diameter at the top surface averaged
over approximately ten experimental runs and averaged over crystal
symmetries.  As the depth of the pack increases the radius of the ring
increases, and interior fills in.  Figure \ref{hcp_rad_dist}(a,b)
shows the radial distribution of intensity for hcp crystals of $9$ and
$25$ layers in depth in response to a quick impulse force over a small
area.  Figure \ref{hcp_rad_dist}(c) shows the position of the peaks in
intensity as a function of the depth of the pack with the error bars
corresponding to the width of the peak at half of the maximum value.
For hcp crystals $25$ layers in depth the intensity is nearly uniform
and thus no lower bound is shown.  Additional data points are included
corresponding to experimental runs with a quick impulse applied to a
large area and to runs with a slowly applied force over a large area.
The sloped line indicates an angle of $27.4$ degrees which is the
average of the angles down the edge and the face of a tetrahedron.
The peaks in intensity and force are found to lie just inside of this
angle as explained in section \ref{Interpretation}.

\subsection{Top surface}

\begin{figure}[p]
\centerline{\epsfxsize= 8.0cm\epsfbox{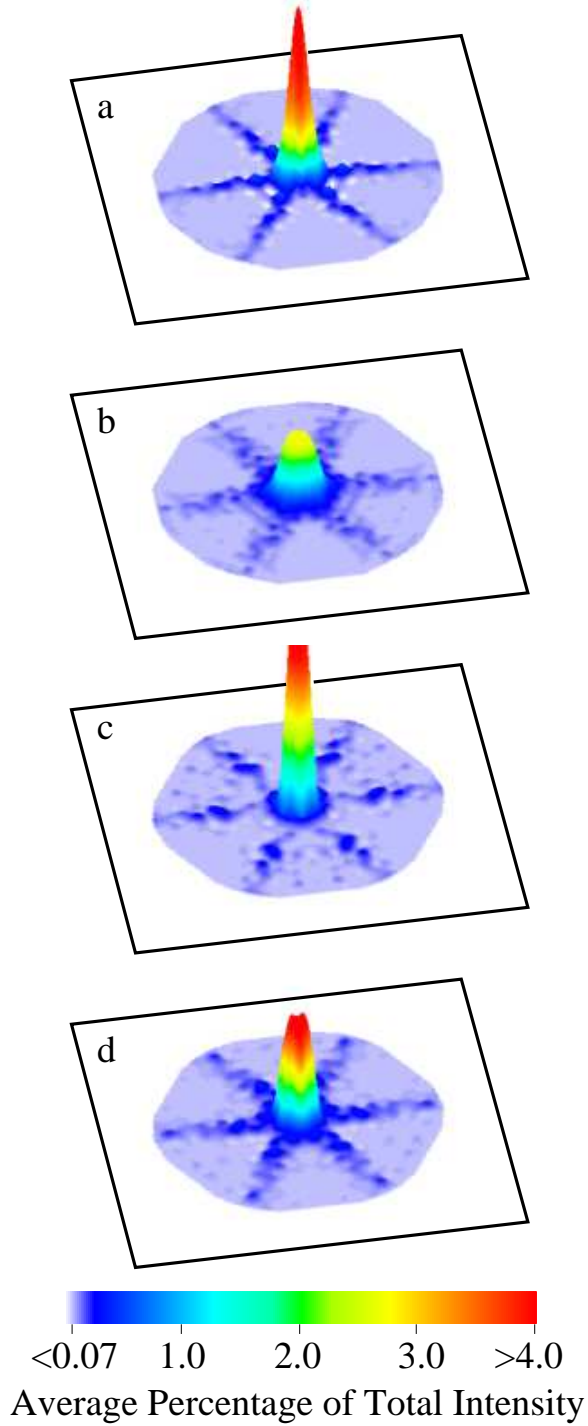}}
\vspace{.5cm}
\caption{(Color) Intensity patterns at the top surface of (a) a $3$
layer hcp crystal, (b) a $9$ layer hcp crystal, (c) a $9$ layer fcc
crystal, and (d) a $19$ layer fcc crystal in response to a quick
impulse force applied to a small area in the center of these images.
Each image is averaged over approximately ten experimental runs and
averaged over crystal symmetries.  The six-fold spoke pattern is
present in both types of crystal packings and does not depend on the
depth of the packing.}
\label{top}
\end{figure}

In addition to measuring the response at the bottom of the crystal
packs, we also used the carbon paper technique to measure the response
at the top of the pack.  Figure \ref{top} shows surface plots for the
average intensity of the carbon dots as a function of position at the
top surface of the crystal in response to a quick impulse applied to
an area two beads in diameter.  Both fcc and hcp crystals show a
characteristic six-fold spoke pattern, which does not depend
significantly on the depth of the crystal.  The magnitude of the
average intensity in these spokes is approximately ten times less than
the intensity in the regions of large force on the bottom surface, but
they are still significantly stronger than the surrounding background.
The spokes extend all of the way to the edge of the analysis region
($15$ bead diameters in radius) without a significant decrease in
intensity.

\section{Geometrical Interpretation}
\label{Interpretation}

\subsection{FCC geometry}

Many aspects of the patterns of force or intensity seen in figures
\ref{force_intensity}, \ref{fcc_surfaces}, \ref{hcp_surfaces}, and
\ref{top} can be explained through force balance in the specific
geometry of the crystals.  Our fcc crystals were oriented with the
$111$ plane horizontal, so that they consisted of a stack of
triangularly ordered planes.  This is equivalent to a typical cannon
ball piling.  In an fcc crystal the stacking order from plane to plane
is such that every third layer lies on top of the first.  This results
in straight lines of contacts between beads running diagonally down
through the pack.

\begin{figure}[p]
\centerline{\epsfxsize= 8.0cm\epsfbox{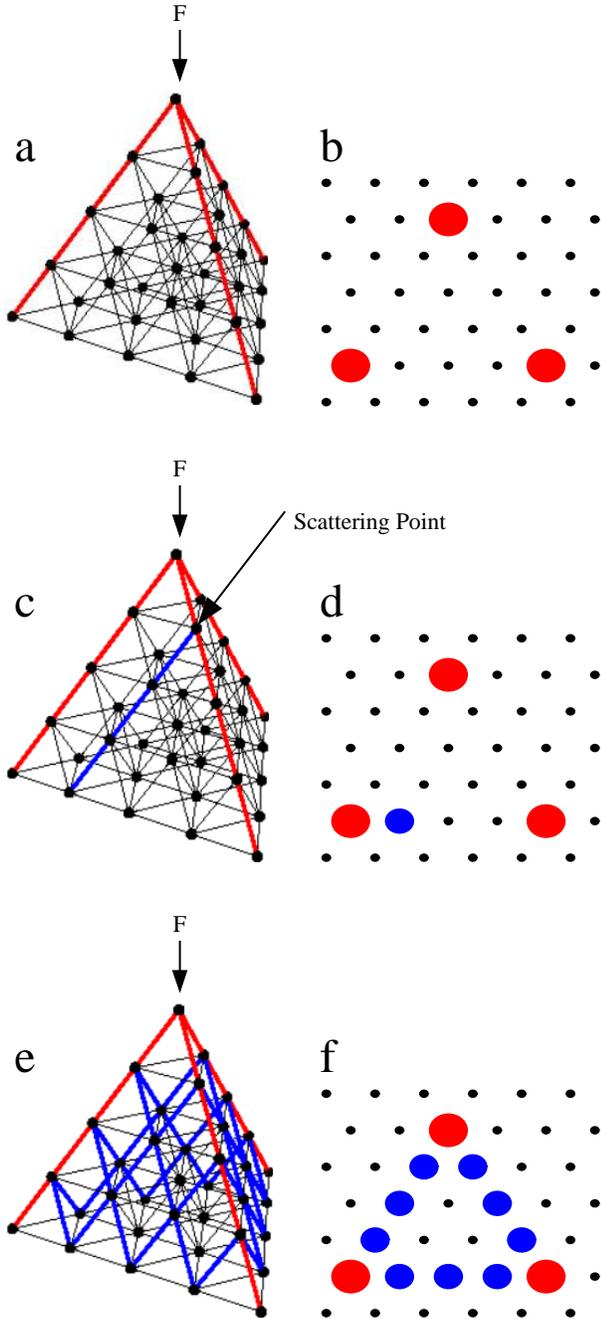}}
\vspace{.5cm}
\caption{(Color) Geometric interpretation of the response to a point
force in an fcc crystal.  (a) In a perfect fcc crystal the force at
the top of the pyramid is balanced by a tripod.  (b) At the bottom
layer only three beads support force.  (c) One point of disorder along
a tripod leg results in a scattering of a small amount of force off of
the tripod leg.  In this case the force is shown being scattered
downward (d)  This scattered force propagates downward to the bottom
surface to one of the lines connecting the regions of large force.
(e) Averaging over all possible points of single downward scattering
events, the scattered forces are contained on the faces of the
pyramidal structure.  (f)  The scattered forces arrive at the bottom
layer arranged along the lines connecting the non-scattered forces.}
\label{fcc_diagrams_1st}
\end{figure}

Figure \ref{fcc_diagrams_1st} shows the geometry of a piece of a
perfect fcc crystal.  If a downward force is applied to the bead at
the top of this pyramidal pile, then the force is transmitted to the
three beads below it in the second layer.  Essentially, the bead at
the top of the pyramid is supported by a tripod, with three lines of
beads stretching downward through the pack.  Only the beads in this
tripod will feel a force in response to the force on the top bead.
All other beads inside and outside of the pyramid are shielded from
that force.  Thus, at the bottom of the crystal there are only three
beads which feel any force.  These beads are arranged in a triangular
pattern with the size of the triangle proportional to the depth of the
pack.  The legs of the tripod in the fcc crystal are at an angle of
$35.3$ degrees with respect to the vertical.  Given this angle the
position of the beads which feel the applied force is determined.  The
solid line in figure \ref{fcc_profiles}(c) shows the predicted angle
of $35.3$ degrees.

\subsection{disorder}

If there is a small amount of disorder in the crystal, then the
positions of the beads within the tripod legs will not be perfectly
aligned.  Thus, in order to balance the forces vectorally, there must
be some force scattered off of the tripod.  For a small degree of
disorder, the majority of the force will still be transmitted along
the tripod legs, with a small amount scattered to one or more
additional neighboring beads.  Figure \ref{fcc_diagrams_1st}(c)
represents an fcc crystal with one point of disorder in the second
layer.  Most of the force is transmitted along the tripod legs, but a
small amount is scattered along the blue line.  If there is no further
disorder, then this scattered force will propagate along its own line
of beads all of the way to the bottom of the pack, and arrive at the
bottom surface along one of the lines connecting the three ends of the
tripod legs, as shown in figure \ref{fcc_diagrams_1st}(d).  Using this
as a model for the effect of one point of disorder we find that, when
considering a small degree of disorder throughout the system, the
majority of the force is transmitted along the tripod legs, with a
small amount scattered down the sides of the pyramid and arriving at
the bottom surface along the edges of the triangle formed by the
tripod legs.  Figure \ref{fcc_diagrams_1st}(e) shows the first order
effect of scattering within an fcc crystal.  The red lines represent
strong force, and the blue lines represent weaker forces resulting
from first order scattering.  The pattern of forces at the bottom of
the crystal, shown in \ref{fcc_diagrams_1st}(f) is seen in the images
in figures \ref{force_intensity} and \ref{fcc_surfaces}.

Disorder can also explain the appearance of the regions of large force
at angles slightly smaller than $35.3$ degrees as seen in figure
\ref{fcc_profiles}(c) as scattered forces arriving preferentially
inside the triangular pattern will skew the regions of large force
towards the inside of the pyramid.  To first order, no force exists
outside of the pyramid.

If at points of disorder the force scatters to beads in plane rather
than to the plane below, then without additional scattering the
scattered force propagates straight outward to the sides of the
container.  For nonclose-packed systems, in which beads in-plane are
not in contact, it is also possible for the forces to scatter upwards
through the pack.  This would produce a spoke-like structure, but the
length of the spokes would be expected to depend on the depth of the
pack which is not seen experimentally.

\begin{figure}[tb]
\centerline{\epsfxsize= 8.0cm\epsfbox{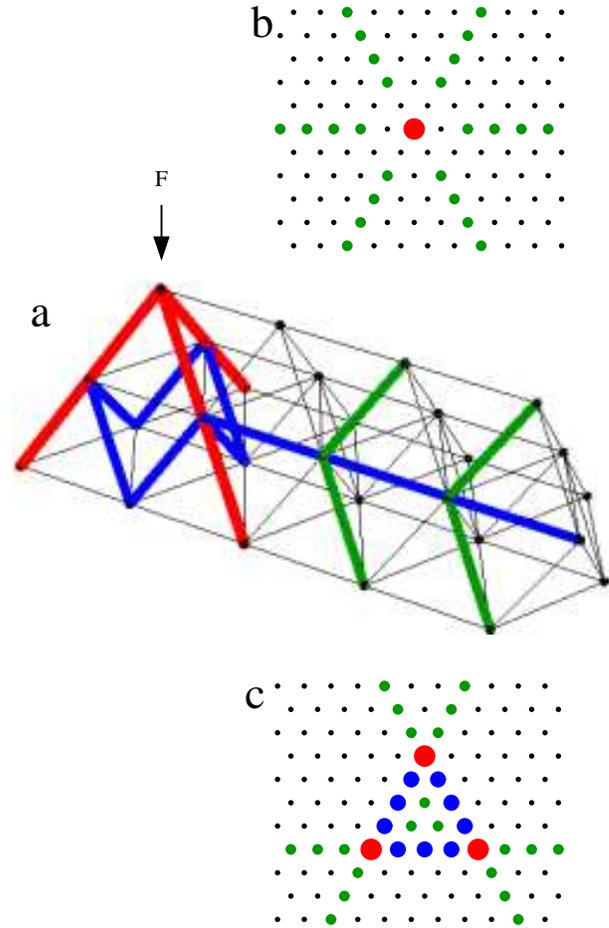}}
\vspace{.5cm}
\caption{(Color) (a) Schematic diagram of an fcc crystal showing
second order scattering.  The red lines represent large unscattered
forces.  Blue lines correspond to first order scattering, both
downward and in-plane, and green lines represent some of the second
order scattering.  Also shown are the patterns at the (b) top and (c)
bottom surfaces of a 5 layer fcc crystal due to second order
scattering.}
\label{fcc_diagrams_2nd}
\end{figure}

The second-order effect of disorder on the transmission of forces is
depicted in figure \ref{fcc_diagrams_2nd}.  Figure
\ref{fcc_diagrams_2nd}(a) shows a piece of a three layer fcc crystal.
The red lines represent the strong forces along the tripod discussed
earlier.  The blue lines are the result of first order scattering with
both in-plane and downward scattering shown.  The green lines
represent some of the scattering from the scattered forces, or second
order scattering.  Of interest is the scattering from the in-plane
forces. These forces can scatter upwards, downwards, or in-plane
through the pack.  Figure \ref{fcc_diagrams_2nd}(b) shows the forces
at the top surface of a five layer deep fcc crystal, where second
order scattering produces a six fold spoke pattern extending radially
outward from the center after a one-bead gap.  These spokes extend
outward to the edge of the system and do not depend on the depth of
the crystal.  Similar experimental spoke patterns are shown in figure
\ref{top}.  Figure \ref{fcc_diagrams_2nd}(c) shows the pattern of
forces at the bottom surface of the crystal where the effect of second
order scattering is to fill in the triangle with very weak forces and
extend the edges of the triangle beyond the vertices. For small
amounts of disorder the magnitude of the forces from second-order
scattering is much smaller than those previously discussed.
Nevertheless figure \ref{fcc_surfaces}(a) shows faint forces along the
lines of the triangle edges beyond the vertices.

\subsection{HCP Geometry}

\begin{figure}[tb]
\centerline{\epsfxsize= 7.0cm\epsfbox{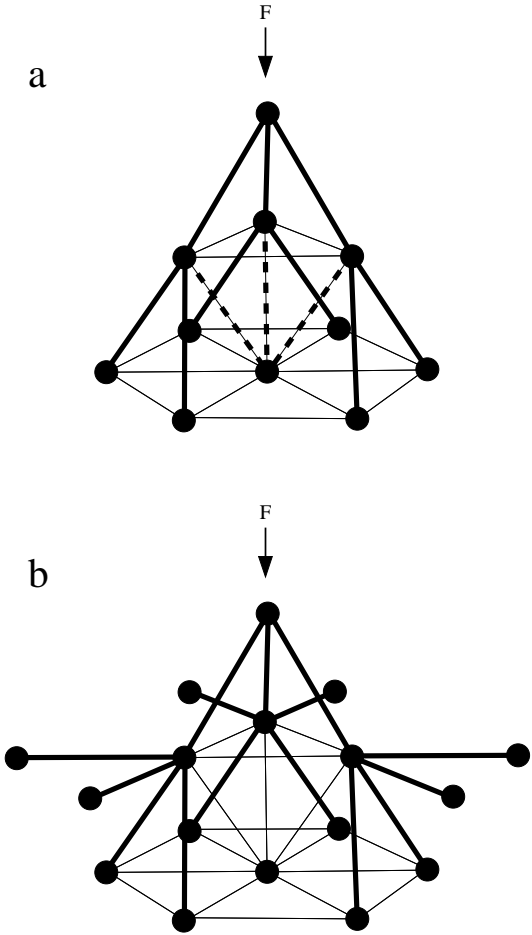}}
\vspace{.5cm}
\caption{Schematic diagrams of a perfect hcp crystal.  (a) Forces
cannot be transfered from the top to the third layer without having
negative forces, shown as dashed lines.  When the weight of the
particles is neglected this pack is unstable without additional
support from the sides.  (b)  Illustration of force being transfered
to in-plane neighbors}
\label{hcp_diagrams}
\end{figure}

The hcp crystals that we studied are also oriented so as to consist of
stacked planes of triangular order.  Unlike the fcc crystals the
stacking of the planes is such that every other plane lies directly
above the first.  The top two layers of an hcp crystal and fcc crystal
are identical, and the force transmission will be the same.  However
the difference in position of the beads in the subsequent layers
causes the force transmission to be markedly different.  In an hcp
crystal there are not straight lines of beads downward through the
pack, so even in a perfect crystal the force cannot be transferred to
a single bead and still preserve vector force balance.  Furthermore,
the nature of the geometry of the hcp crystal is such that the force
on a bead in the second layer received from the top bead cannot be
transferred to the three beads below.  This would require a negative
force (see figure \ref{hcp_diagrams}) or adhesion, which is not
present in dry granular systems.  The result is that a pack such as
that shown in figure \ref{hcp_diagrams}(a) is not stable to an
external force when there is no friction and the weight of the
individual grains is neglected.  If only the weight of the beads is
considered then a perfect hcp packing is stable down to five layers
after which it develops the same instability.  In order to keep the
beads from moving there must be additional support.  Inside a larger
crystal this instability demands that forces are transmitted to beads
in-plane for close-packed crystals and that forces are transmitted
upwards for nonclose-packed crystals.  If forces are allowed to be
transmitted to neighbors in-plane then figure \ref{hcp_diagrams}(b)
shows the propagation of forces for a small piece of a perfect hcp
crystal in response to a downwards force at the top.  In-plane forces
will continue directly to the walls without additional scattering in a
perfect crystal.  First order scattering will result in a spoke like
pattern at the top surface similar to that seen in the fcc crystals.
The expected pattern at the bottom surface is a ring of maximum force,
whose radius grows with additional layers of depth at angles of $19.5$
degrees and $35.3$ degrees for alternating layers.  The solid black
line in figure \ref{hcp_rad_dist} shows the expected radius of the
ring as a function of depth based on an average angle of $27.4$
degrees.  As the depth of the pack increases the interior of the ring
also fills in.  This process occurs even in a perfect hcp crystal.
Disorder has little effect on the pattern of forces on the bottom
surface, but does result in a spoke pattern on the top surface, as is
seen experimentally in figure \ref{top}.

\section{conclusions}

We have observed that in the highly ordered regime the response to a
localized force is greatly dependent on the structure of the beads
within a granular pack.  We found distinctly different responses from
fcc crystals and hcp crystals, both of which behave differently than
amorphous packs.  Amorphous packs are found to respond with a large
central peak.  fcc crystals respond with three regions of large force
at the bottom surface, while hcp crystals show rings of maximum force.
These structures can be explained by force balance in the geometry of
the crystal.  These findings are consistent with the two-dimensional
results of Geng {\em et al.} \cite{Geng00,Behringer_Geng_talks}.

The texture of a granular pack is crucially important for determining
the average pattern of stress transmission, as has been suggested by
Rajchenbach \cite{Rajchenbach}.  The fcc crystal structure, without
disorder, leads to a tripod of force chains that is the dominant
feature of the force profiles.  Small amounts of disorder create
secondary structures at the bottom surface and are responsible for a
six-fold spoke pattern of forces at the top surface.  The hcp crystal
structure does not allow for straight line propagation of forces
downward through the pack.  Even without disorder the forces must
split at every layer.

The splitting of forces within an hcp pack can be viewed in analogy to
the forcechain splitting model of Bouchaud {\em et al.}
\cite{Bouchaud00}.  In a perfect hcp crystal the force splitting is
driven by the orientations of neighboring grains and is thus nearly
everywhere uniform, in constrast to the forcechain splitting model in
which randomly placed defects drive forces to split into random
angles.  Nevertheless, the average response of the system to multiple
splittings is found to have similarities across these two systems.
The forcechain splitting model has been shown to lead to a central
minimum of force at the bottom surface in response to a localized
force at the top for small packing heights, which is consistent with
our observations on hcp crystals of small heights.  In two-dimensional
simulations of forcechain splitting with a small density of defects
the response pattern made a transition towards elastic-like behavior
with a broad central peak for depths larger than approximately six
mean free paths of the force \cite{Bouchaud00}.  We find that for
three-dimensional hcp crystals larger than approximately ten layers in
depth the central minimum of force fills in leading to a broad central
peak response.  This experimental result in three dimensions with
splitting generated uniformly throughout the pack by contact
orientations agrees qualitatively with the predictions of the
forcechain splitting model simulations in two dimensions where
randomly placed defects generate the splitting.

Finally, it is important to note that previous experiments have shown that the
crystal structure does not affect $P(F)$, the probability distribution of
interparticle contact forces \cite{Blair01}.  We conclude that the
fluctuations of forces within a single pack are dominated by microscopic
details of the granular pack and are not influenced by the arrangement of
the beads.  Nevertheless, the average response to a localized force is
greatly influenced by this structure.

\section{Acknowledgments}
We would like to thank Adam Marshall, Daniel Blair, Daniel Mueth, and
Milica Medved for their assistance with this project.  This work was
supported by NSF under Grant No.\ cts0090490 and by the MRSEC Program
of the NSF under Grant No.\ DMR-9808595.

\end{document}